\documentclass[conference]{IEEEtran}
\IEEEoverridecommandlockouts
\makeatletter
\renewcommand{\fnum@figure}{Fig. \thefigure}
\makeatother
\usepackage[left=1.608cm,right=1.608cm,top=1.8cm]{geometry}
\usepackage{comment}
\usepackage[noadjust]{cite}
\usepackage{filecontents}
\usepackage{balance}

\usepackage{graphicx}
\usepackage{epstopdf}
\usepackage{amsfonts}
\DeclareGraphicsExtensions{.pdf,.jpeg,.png,.eps}
\usepackage[cmex10]{amsmath}
\usepackage{mathabx}
\usepackage{algorithmic}
\usepackage{array}
\usepackage{mdwmath}
\usepackage{mdwtab}
\usepackage{eqparbox}
\usepackage{url}
\usepackage{color,soul}
\usepackage{amsmath} 
\usepackage{multirow} 

\usepackage{dirtytalk} 

\usepackage[utf8]{inputenc}
\usepackage[english]{babel}
\usepackage{fancyhdr}
\usepackage{lastpage}
\usepackage{caption}
\usepackage{subcaption}
\usepackage{adjustbox}
\usepackage{amssymb}
\usepackage[ruled,vlined]{algorithm2e}
\usepackage{xcolor}

\usepackage{calrsfs}
\DeclareMathAlphabet{\pazocal}{OMS}{zplm}{m}{n}

\usepackage[acronym]{glossaries}
\newacronym{IDS}{IDS}{intrusion detection system}
\newacronym{CNN}{CNN}{convolutional neural network}
\newacronym{LSTM}{LSTM}{long short-term memory}
\newacronym{AI}{AI}{artificial intelligence}
\newacronym{DL}{DL}{deep learning}
\newacronym{ML}{ML}{machine learning}
\newacronym{FP}{FP}{false positive}
\newacronym{FN}{FN}{false negative}
\newacronym{TP}{TP}{true positive}
\newacronym{TN}{TN}{true negative}
\newacronym{UAVs}{UAVs}{unmanned aerial vehicles}
\newacronym{UAV-IDS}{UAV-IDS}{unmanned aerial vehicle intrusion detection system}
\newacronym{PCA}{PCA} {principal component analysis}
\newacronym{Relu}{Relu}{rectified linear unit}
\newacronym{IOT}{IOT}{internet of things}
\newacronym{NIDS}{NIDS}{network intrusion detection system}
\newacronym{HIDS}{HIDS}{host-based IDS}
\newacronym{1D}{1D}{one dimensional}
\newacronym{RF}{RF}{radio frequency}
\newacronym{DNN}{DNN}{deep neural networks}
\def\BibTeX{{\rm B\kern-.05em{\sc i\kern-.025em b}\kern-.08em
    T\kern-.1667em\lower.7ex\hbox{E}\kern-.125emX}}

\begin{document}

\bstctlcite{IEEEexample:BSTcontrol}
\title{Enhanced Intrusion Detection System for Multiclass Classification in UAV Networks\\
}

\author{\IEEEauthorblockN{Safaa Menssouri\IEEEauthorrefmark{1},
 Mamady Delamou\IEEEauthorrefmark{1}, Khalil Ibrahimi\IEEEauthorrefmark{2}, El Mehdi Amhoud\IEEEauthorrefmark{1}
}
\IEEEauthorblockA{\IEEEauthorrefmark{1}College of computing, Mohammed VI Polytechnic University, Ben Guerir, Morocco \\
\IEEEauthorrefmark{2}Ibn Tofail University, Faculty of Sciences, Laboratory of Research in Informatics (LaRI), Kenitra, Morocco
} 
{\{safaa.menssouri, mamady.delamou, elmehdi.amhoud\}@um6p.ma}, ibrahimi.khalil@uit.ac.ma}

\maketitle
\begin{abstract}
Unmanned Aerial Vehicles (UAVs) have become increasingly popular in various applications, especially with the emergence of 6G systems and networks. However, their widespread adoption has also led to concerns regarding security vulnerabilities, making the development of reliable intrusion detection systems (IDS) essential for ensuring UAVs safety and mission success. This paper presents a new IDS for UAV networks. A binary-tuple representation was used for encoding class labels, along with a deep learning-based approach employed for classification. The proposed system enhances the intrusion detection by capturing complex class relationships and temporal network patterns. Moreover, a cross-correlation study between common features of different UAVs was conducted to discard correlated features that might mislead the classification of the proposed IDS.
The full study was carried out using the UAV-IDS-2020 dataset, and we assessed the performance of the proposed IDS using different evaluation metrics. The experimental results highlighted the effectiveness of the proposed multiclass classifier model with an accuracy of 95\%.
\end{abstract}
\section{Introduction And Related Works}
Due to dynamic reconfigurability, fast response, and ease of use, \gls{UAVs} commonly called drones have gained popularity in many diverse applications, including delivering medicines and goods, military surveillance, agriculture, and first aid \cite{Applications}. The pervasiveness of \gls{UAVs}, as well as the invasive deployment of this technology in several key domains, has raised security issues, particularly in the context of networked UAVs within the emerging landscape of 6G systems. Over open-air communication channels, networked UAVs are vulnerable to a variety of malicious assaults.\\
\indent While a UAV system is made up of a network of \gls{UAVs} and a ground control center (GCC), cyber-attacks can target any system component. As a result, \gls{IDS} has become an important process in network security that aims to monitor the network from abnormal activities.\\ 
\indent Although there are different types of \gls{IDS} \cite{choudhary2018intrusion}, we are only interested in anomaly-based \gls{IDS}. These systems attempt to detect known and unknown attacks providing a more comprehensive detection capability than other \gls{IDS} including signature-based, which only detects attacks based on pre-defined known signatures \cite{10000726,s22218085}.
In fact, various methods to detect and identify intruding drones have been discussed in the literature such as radar video, acoustic sensors \cite{8102043}, Wi-Fi sniffing \cite{8556480}, and \gls{RF} sensing.\\
\indent In \cite{8846214}, the authors illustrated the concept of drone detection and classification  using \gls{ML} techniques. Essentially, the application of \gls{ML} helps to detect drones as either drone or attack in a binary classification model. However, some studies in the literature go beyond traditional classifications and conduct multiclass classification to identify drone types.
With this in mind, the authors of \cite{ALSAD201986} presented an open-source RF drone dataset and system capable of not only detecting but also identifying drones based on RF communication between drones and their controllers. The interesting here is that the authors have adopted \gls{DL} algorithms to detect and classify drones, contrary to the common trend of relying on traditional \gls{ML} approaches. However, the authors also report that the algorithm failed to identify two drones made by the same company and their modes. Hence, a solution with different DL architectures such as \gls{CNN} has been implemented to solve this issue in another study conducted by the authors of \cite{9089489}. The findings proved the effectiveness of using \gls{CNN} model in order to detect drones with an accuracy over 99\%. \\
\indent Given the fact that UAVs are commonly used in civilian environments, traditional physical detection techniques such as radar, vision, and sound may be inadequate in many situations. To address these challenges, leveraging encrypted Wi-Fi traffic data from UAVs holds great promise for detecting unauthorized UAV intrusions. Therefore, the authors of \cite{10.1007/s00521-022-07015-9} used a \gls{DL} approach with the \gls{CNN}, and the proposed model considers encrypted Wi-Fi traffic data records from the UAV-IDS-2020 dataset. However, their experimental results were limited to the binary classifier.\\
\indent In this paper we propose a principal component analysis, one-dimensional convolutional neural networks, and long short-term memory (PCA-1D CNN-LSTM) approach to enhance UAV network intrusion detection efficiency. The main contributions of this paper are summarized as follows:

\begin{table*}[t]
  \caption{UAV-IDS-2020 dataset distribution}
  \centering
  \begin{tabular}{ccccccccc}
    \hline
    \multirow{2}{*}{\textbf{Data distribution}} & \multicolumn{2}{c}{\textbf{Parrot\_UAV}} & \multicolumn{2}{c}{\textbf{DBP\_UAV}} & \multicolumn{2}{c}{\textbf{DJI spark}} & \multicolumn{2}{c}{\textbf{All UAVs}}\\
     & Uni-Dir &  Bi-Dir   &  Uni-Dir &  Bi-Dir  &   Uni-Dir &  Bi-Dir  &   Uni-Dir &  Bi-Dir \\
    \hline
    Total no. of samples & 11663 & 19380 & 14864 & 17256 & 73 & 5500 & 26600 & 42136\\
    No. of training samples & 1063 & 1751 & 1351 & 1569 & 7 & 500 & 2421 & 3820 \\
    No. of testing samples & 10600 & 17629 & 13513 & 15687 & 66 & 5000 & 24179 & 38316 \\
    Train-Test Percent  & 10-90\% & \multicolumn{5}{c}{Overall Total} & 68736 & \\
    \hline
  \end{tabular}
  \label{tab: dataset distribution}

\end{table*}

\begin{itemize}
\item[$\bullet$]To the best of our knowledge, we are the first to perform a multi-class classification on the recent datasets for networked \gls{UAVs} (UAV-IDS-2020).
\item[$\bullet$] We design an IDS that employs a binary-tuple representation for encoding class labels along with a deep learning-based approach to improve overall system performance.
\item[$\bullet$] While our developed  architecture is capable of achieving almost 100\% accuracy in binary classification tasks, we leverage the cross-correlation between common features along with a \gls{PCA} based dimensionality reduction method for the multiclass task.
\end{itemize}

 The remainder of the paper is organized as follows: In Section II, we detail the architecture of our proposed deep learning-based model along with a description of the dataset. In Section III, we present simulation results, we discuss the findings, and we also provide a comparative study. Finally, in Section IV, we conclude and outline our perspectives.

\section{System Model And Problem Formulation}
\begin{figure}[t]
\centering    \includegraphics[width=0.32\textwidth]{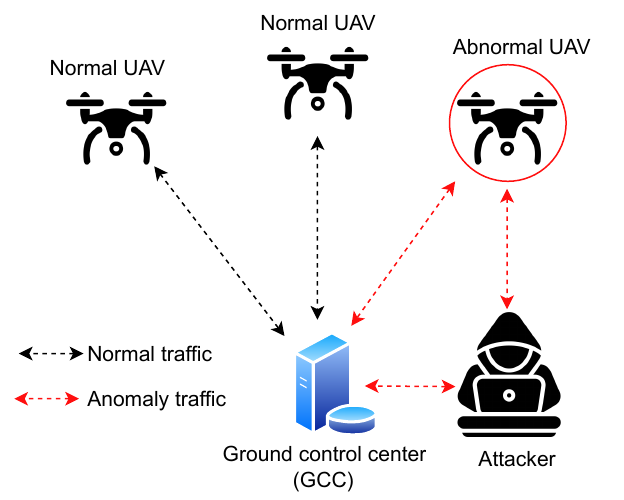}
    \caption{System model.}
    \label{fig: system Model}
 \vspace{-1.5em}
\end{figure}

\indent Fig. \ref{fig: system Model} illustrates our system model. We have a scenario that features a network comprising both normal \gls{UAVs} and abnormal ones, all engaged in communication with a GCC. These \gls{UAVs} belong to three distinct UAV types. Within this setup, UAVs are considered abnormal when an attacker takes control of an existing UAV within our network via the GCC or injects an entirely new unauthorized UAV into the system. Our aim is therefore to develop a deep learning-based system that can distinguish between normal and abnormal UAVs, and also identify the specific type of the drone involved.

 \subsection{UAV-IDS Dataset}
In our study, we used the open-source UAV-IDS-2020 cybersecurity dataset \cite{10.1145/3219819.3220117}. This dataset comprises up-to-date and unique entries of encrypted Wi-Fi traffic concerning three types of \gls{UAVs}: Parrot Bebop, DBPower UDI, and DJI Spark. Each entry is accompanied by binary labels indicating whether it corresponds to a drone intruder (labeled as 1) or not (labeled as 0). Alongside the training and testing datasets, the dataset furnishes supplementary meta-information that helps to perform additional \gls{ML} tasks. For example, it contains the computational generation time for each statistical attribute, as well as information about the computational dependency among different features through the incidence matrix.\\
\indent Moreover, the dataset is available for two modes of UAV communication flow: bidirectional flow (BF), which comprises the uplink flow, downlink flow, and total traffic; and unidirectional flow (UF), which includes only the total traffic flow. For each type of flow, the raw data sources are the packet sizes and packet inter-arrival time, and for each source, the nine statistical measures including the $mean$, $median$, $mad$, $max$, $min$, $std$, $mean~square$, $kurtosis$, and $skewness$ are extracted. Thus, for the bidirectional flow, there are 9 features × 2 sources × 3 direction flow = 54 features (+1 label) whereas there are 9 features × 2 sources = 18 features (+1 label) for the unidirectional flow dataset.
The dataset consists of 68,736 examples in total. To gain more insight, Table \ref{tab: dataset distribution} summarizes the traffic distribution of the UAV-IDS-2020 dataset.

 \subsection{Principal Component Analysis (PCA)}
In the field of data analysis, \gls{PCA} is a dimensionality reduction technique \cite{9264568}. 
In our study, we employed \gls{PCA} as a crucial preprocessing step in our multiclass classification task. We chose \gls{PCA} to deal with the high dimensionality inherent in the dataset, which comprises a substantial number of features extracted from UAV communication flow attributes. By transforming the original feature space, \gls{PCA} enabled us to capture the most significant variations in the dataset, thereby improving performance, reducing training time, and minimizing the risk of overfitting.

\subsection{PCA-1D CNN-LSTM System}
\begin{figure}[t]
\centering
\includegraphics[width=0.48\textwidth]{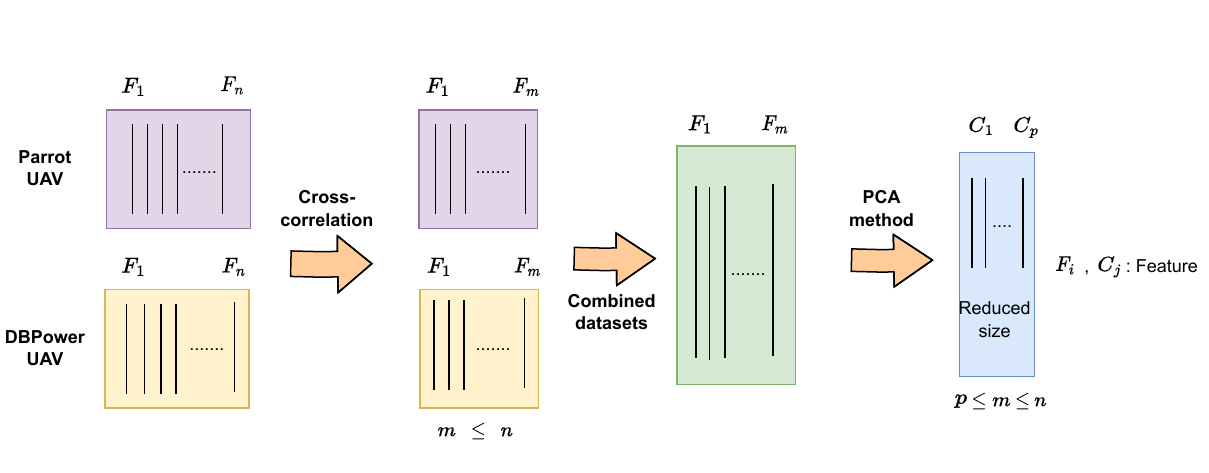}
    \caption{Preprocessing diagram where $m$, $n$, and $p$ are the number of features in the datasets.}
    \vspace{-0.5em}
    \label{fig: Preprocessing Diagram}
\end{figure}

\begin{figure*}[t]
\centering
    \includegraphics[width=0.63\textwidth]{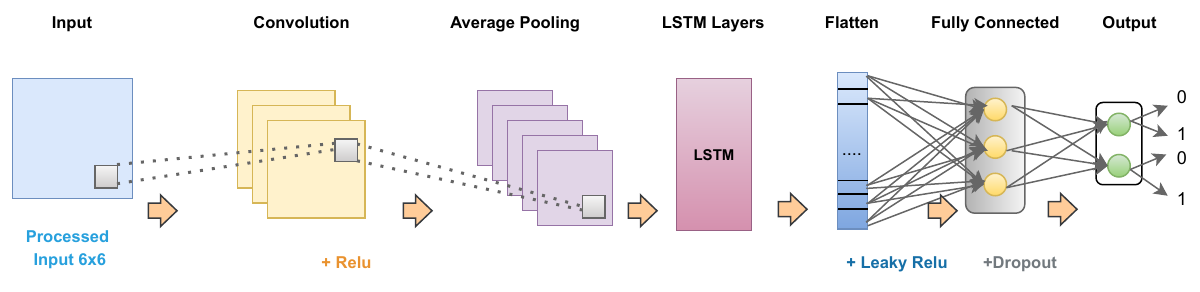}
    \caption{Structure of the proposed 1D CNN-LSTM model.}
        \vspace{-1em}
    \label{fig: Model Architecture}
\end{figure*}

The \gls{UAV-IDS} workflow we followed comprises several sequential steps. It begins with the importation of the dataset, catering to both binary and multiclass classification tasks. Specifically, we acquired datasets for the two communication flow modes, unidirectional and bidirectional, for each of the three UAV types. In addition, for binary classification, we created a combined dataset by concatenating data from all three UAV types within each communication flow mode. However, for multiclass tasks, due to a lack of data for the DJI Spark type in the unidirectional flow mode, we concatenated data from only two UAV types to create the combined dataset.\\
\indent Subsequently, the dataset undergoes preprocessing, encompassing tasks such as data labeling/encoding, data generation, performing the cross-correlation between common features and \gls{PCA} method for the multiclass classification as depicted in Fig. \ref{fig: Preprocessing Diagram}. Data padding, and reshaping are also part of the preprocessing steps. 
\\ In the following, we denote by \( M = \{m_1, m_2,..., m_p\}  \in \mathbb{R}^{n \times p} \) the input matrix, where $n$ and $p$ represent respectively the number of samples and features in our dataset. Each \( m_i  \in \mathbb{R}^{n} \) represent the vector of values corresponding to the network traffic feature $M_{i}$, \(\forall i \in \{1,..., p\} \).
\\ Let $M$ and $M'$ be the input matrices representing the dataset of UAV-1 and UAV-2 respectively:
\begin{enumerate}
    \item Calculate the correlation coefficient \(Cor(m_i, m'_i)\) between common features $M_{i}$ and $M'_{i}$ of both UAVs for $1\le i \le p$ as follows:
    \begin{equation}
    Cor(m_i, m'_i) = \frac{Cov(m_i, m'_i)}{\sigma_{m_i} \sigma_{m'_i}},
    \end{equation}
with \( \sigma_{m_i} \) and \( \sigma_{m'_i} \) being the standard deviation of \( m_i \) and \( m'_i \) respectively, and \(Cov(m_i, m'_i)\) being the covariance between $m_{i}$ and $m'_{i}$.
\item Drop highly correlated features, and retain only those  having a correlation value below a certain threshold \( \theta \). We denote by $M_{f}$ the set of retained features such that:
\begin{equation}
    M_{\text{f}} = \{ M_i \mid \forall i \in \{1,\ldots, p\},\text{ } Cor(m_i, m'_i) \leq \theta \}.
\end{equation}

    \item Standardize the selected features $M_{f}$, and then apply the PCA reduction method to the resulting data.
    \end{enumerate}
 \indent In the next phase, we trained and validated the proposed model combining 1D \gls{CNN} and \gls{LSTM} architectures using 10\% of the dataset (Table \ref{tab: dataset distribution}). This training process employs a five-fold cross-validation approach. Once trained, we evaluated the performance of the proposed model using the remaining (unseen) 90\% of the dataset (Table \ref{tab: dataset distribution}), employing diverse evaluation metrics including accuracy, loss, and confusion matrix. For the binary classification, the system's output distinguishes between normal UAV and intruder (attack) instances. Furthermore, for the multiclass task, the system also identifies the specific type of the drone involved. The structure of the proposed model is presented in Fig. \ref{fig: Model Architecture}.   
 \\ \indent We opted for a combination of CNN and LSTM in our model due to the essential roles they play in achieving our classification objectives. The LSTM's ability to capture temporal dependencies is crucial for analyzing packet inter-arrival times features, while the CNN's feature extraction and classification capabilities make it highly effective for discerning complex patterns, particularly in classifying drones and their types.

\subsection{Performance Evaluation Metrics}
To evaluate the performance of our proposed model, we used specific metrics, including accuracy ($AC$), precision ($PR$), error ($Er$), detection rate ($DR$), false alarm rate ($FAR$), and $F1-score$ via confusion matrix, along with the loss ($L$). These performance metrics are defined as follows:
\begin{equation}
AC=\frac{TP+TN}{TP+TN+FP+FN}, \mathrm{~} PR=\frac{TP}{TP+FP},
\end{equation}
\begin{equation}
DR=\frac{TP}{TP+FN}, \mathrm{~} FAR=\frac{FP}{FP+TN},
\end{equation}
\begin{equation}
F1-score=\frac{2TP}{2TP+FP+FN}, \mathrm{~}  Er = 1-AC.
\end{equation}
Where $TP$, $TN$, $FP$ and $FN$ represent true positives, true negatives, false positives, and false negatives, respectively.\\
\indent The following binary cross-entropy loss function is used to train the model:
\begin{equation}
L= -\frac{1}{2N} \sum_{i=1}^{N} \sum_{j=1}^{2} y_{ij} \log(p_{ij})+(1 - y_{ij}) \log(1 - p_{ij}),
\end{equation}
where $N$ is the number of samples, $y_{ij} \in \{0,1\}$ is the true label of the $j$th element of the $i$th sample output, and $p_{ij}$ is the predicted probability of the $j$th element of the $i$th sample output. For binary classification, $j=1$ and for multiclass classification $1\le j \le2$. Furthermore, the total loss $L_{T}$ we used to evaluate our model can be expressed as $L_T = \frac{1}{K} \sum_{k=1}^{K} L_k$, with \( K \) representing the number of folds, and \( L_k \) denoting the loss obtained after the \( k \)-\( th \) fold in a cross-validation setting.
\section{Experimental Results}
In this section, we investigate and evaluate the performance of the \gls{DL} based architecture proposed in Section II-D using the designed binary and multiclass approaches.  To implement these experiments, we have tuned the hyperparameters to find the suitable ones (which are presented in Table \ref{tab: Parameters of the System}) that achieve the optimal performance of the proposed model.  In addition, to enhance the performance of the proposed model, we applied K-fold cross-validation. Specifically, a five-fold cross-validation was used, which divides randomly the dataset into five sections. In each round, one section was used for testing, keeping the rest four sections for training. Moreover, the number of training epochs is set to 100 for our 1D CNN-LSTM model, with the choice of the stochastic gradient descent (SGD) with momentum as the optimizer due to its computational efficiency. Finally, we added the dropout layer with a rate value equal to 0.4 in the model to improve generalization performance and reduce overfitting.\\The weight update rule using the SGD optimizer is as follows:
\begin{equation}
v_{t} = \beta v_{t-1} + (1 - \beta) \nabla_{W} L(W_{t}), 
\end{equation}
\begin{equation}
W_{t+1} = W_{t} - \alpha v_{t}, 
\end{equation}
where \(W_{t}\) represents the weight matrix at time step \(t\), \(\alpha\) is the learning rate, \(\nabla_{W} L(W_{t})\) represents the gradient of the loss function with respect to the weights, \(\beta\) is the momentum coefficient, and $v_{t}$ is the momentum term at time step $t$.
 
\subsection{Binary Classification}
\begin{figure*}[t]
\centering
     \begin{subfigure}[b]{0.31\textwidth}
         \includegraphics[width=\textwidth]{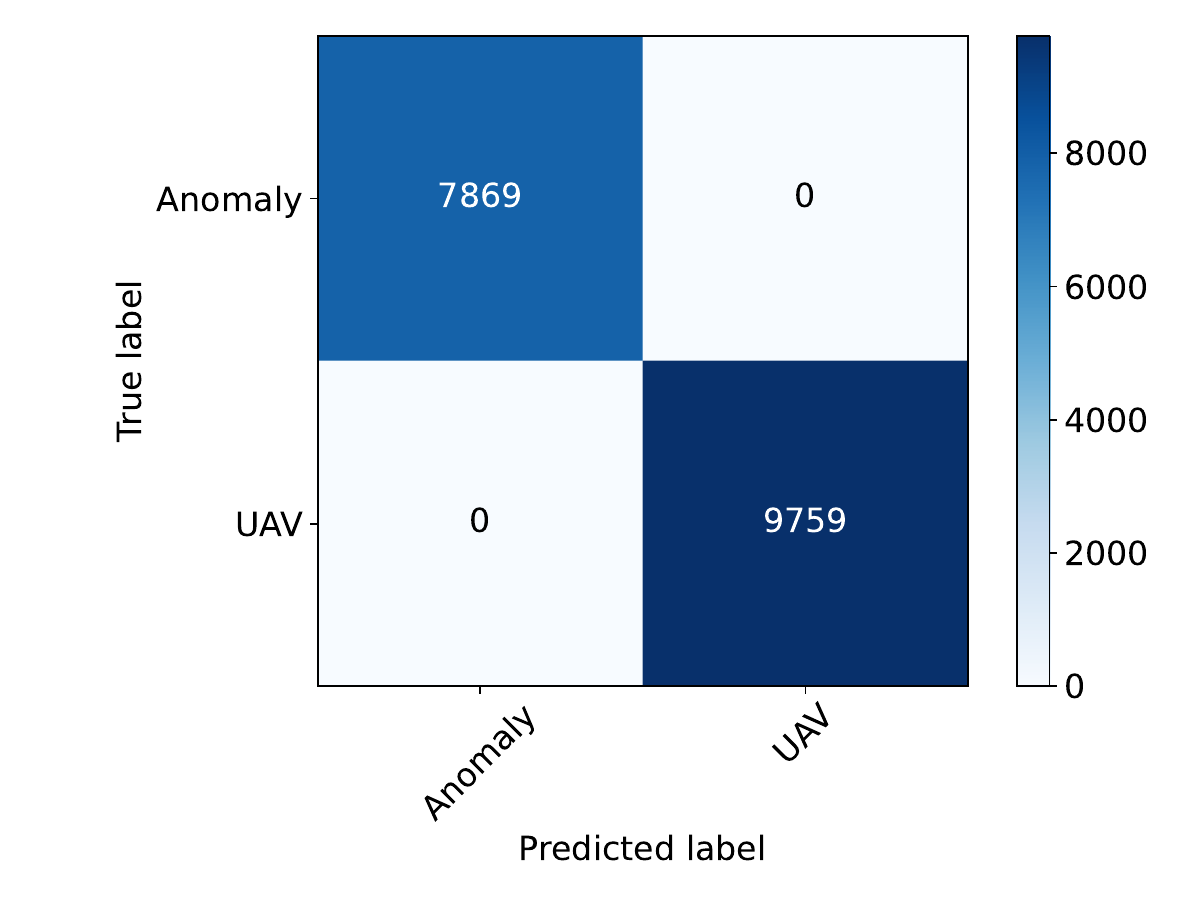}
         \caption{Confusion matrix of parrot UAV (BF).}
         \label{fig: Confusion matrix for the parrot BF}
     \end{subfigure}
          \begin{subfigure}[b]{0.34\textwidth}
         \includegraphics[width=\textwidth]
         {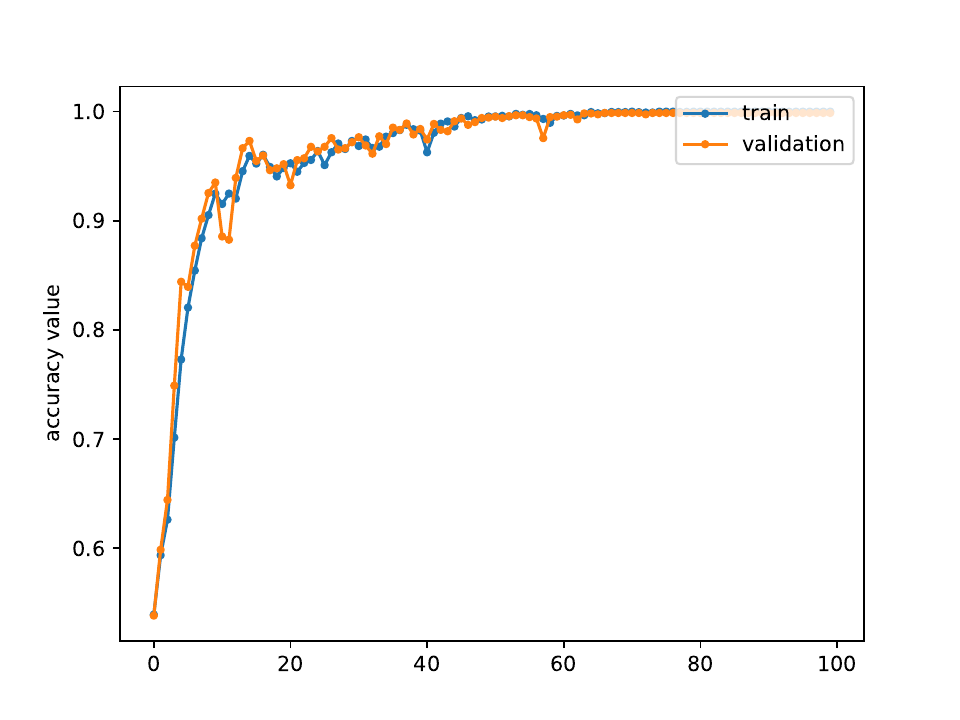}
         \caption{Average accuracy for the combined dataset (UF).}
         \label{fig: Accuracy for the Combine UF}
     \end{subfigure}
     \begin{subfigure}[b]{0.31\textwidth}
         \includegraphics[width=\textwidth]{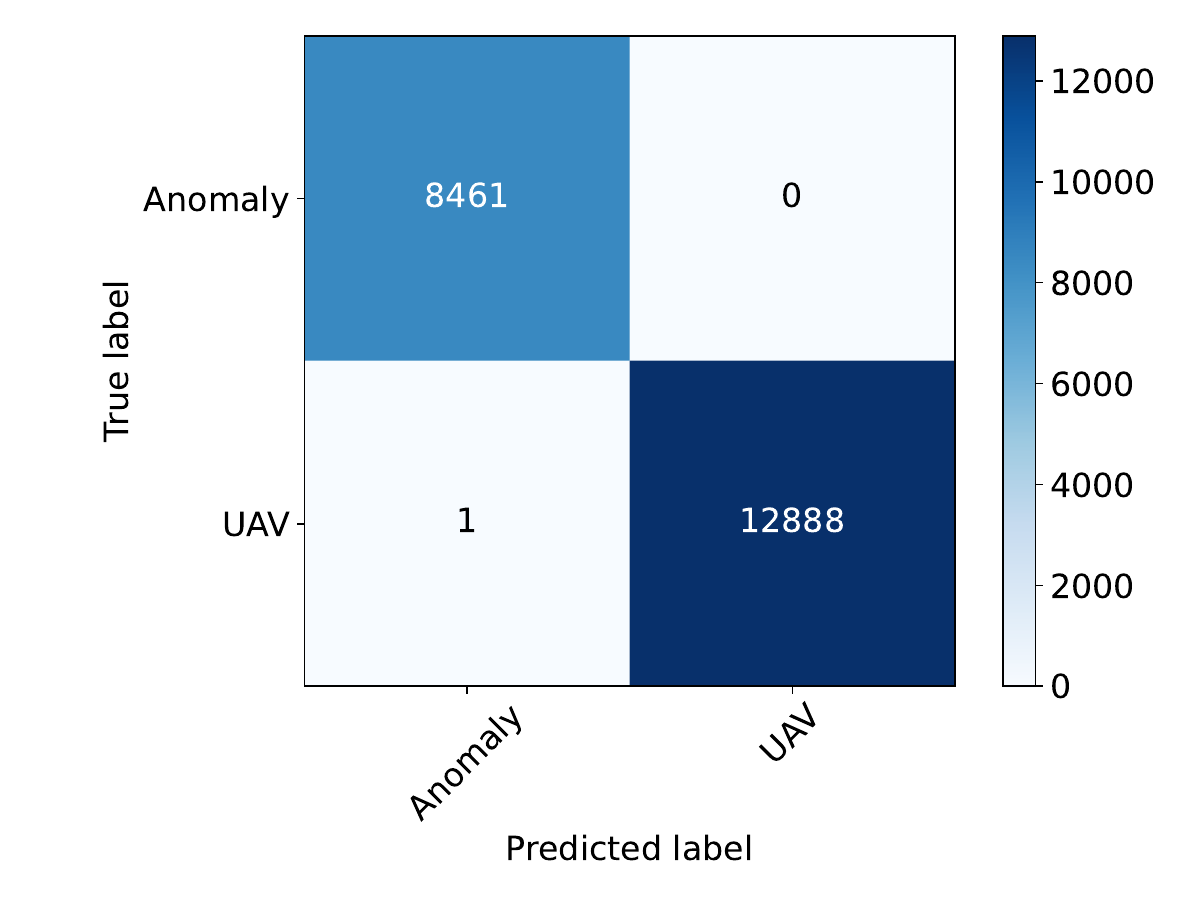}
         \caption{Confusion matrix for the combined (UF).}
         \label{fig: Confusion matrix for the Combine UF}
     \end{subfigure}
        \caption{The average accuracy and confusion matrix for the two-class model for parrot type in bidirectional flow (BF) and the combined dataset in unidirectional flow mode (UF).}
        \vspace{-1.3em}
        \label{fig: Accuracy and Confusion matrix for the binary classifier model for parrot BF and Combine UF}
\end{figure*}

We performed the binary classification on the UAV-IDS-2020 dataset using our model 1D CNN-LSTM with an input of size 8x8 as the best size after testing different sizes. This input size was achieved by adding zero-padding to the dataset during the preprocessing step. Table \ref{tab: Accuracy comparaison} shows the results of these experiments for the combined dataset.\\
\indent The confusion matrices for the binary classifier model for parrot type in BF and for the combined dataset in UF mode are represented in Fig. \ref{fig: Confusion matrix for the parrot BF} and Fig. \ref{fig: Confusion matrix for the Combine UF} respectively. The obtained results show the high performance of our proposed model in differentiating between normal UAV and anomaly classes when applied to unseen data, with almost 100\% accuracy and 0\% loss in most cases. Furthermore, when we consider other evaluation metrics such as precision and F1-score, our model consistently achieves an impressive 99.99\% performance level. In addition, from Fig. \ref{fig: Accuracy for the Combine UF} which represents the average accuracy for the combined dataset in UF mode, we can conclude that the accuracy achieves 100\% after around 40 epochs.

We compared our model with the state-of-the-art intrusion detection methods on the UAV-IDS-2020 dataset. By using the same experimental settings, our model's performance surpassed existing methods in terms of accuracy for Parrot and DBPower UAVs, as well as for the combined dataset (Comb.) notably in UF, with 99.99\%, 99.90\%, and 99.98\% accuracy rate respectively. As depicted in Table \ref{tab: Comparaison Model}, our model outperforms the UAV-IDS-ConvNet proposed by the authors of [10]. This comparison underscores the efficacy of our approach, especially in scenarios where the number of samples is limited (Table \ref{tab: dataset distribution}). Moreover, it fully proves that our proposed model can improve the detection accuracy and has good generalization ability for the detection of unseen datasets.

\begin{table}[t]
  \caption{Relevant parameters of the proposed system}
    \centering
    \begin{tabular}{c|c|c}
    \hline
       \textbf{Model or Method} & \textbf{Parameters} &  \textbf{Values}\\
       \hline
                   PCA & Scaler & StandardScaler\\
                       & Num\_components & 12 \\
        \hline
                       & Convolution & Filter=5, Kernel\_size=4×4\\
                       & Average Pooling & Pool\_size=2×2\\
                       & Num\_units(LSTM) & 32\\
                       & Dense (FC) & 100\\
              CNN-LSTM & Dense & 2\\
                       & Optimizer & SGD\\
                       & Activation function & Sigmoid\\
                       & Epoch / Batch\_size & 100 / 50\\   
       \hline
    \end{tabular}
    \label{tab: Parameters of the System}
\end{table}

 \begin{table}[t]
 \centering
  \caption{Accuracy (\%) comparison for different input sizes of combined dataset }
    \centering
    \begin{tabular}{|c|c|c|c|c|}
    \hline
    \textbf{Input size} & \textbf{28 * 28} &  \textbf{20 * 20} &  \textbf{8 * 8} &  \textbf{9 * 6}\\
       \hline
          Combined UF & 85.65 & 89.97 & 99.98 & 90.70 \\
        \hline
          Combined BF & 99.98 & 99.99 & 99.99 & 99.98 \\ 
       \hline
    \end{tabular}
    \vspace{-1.5em}
    \label{tab: Accuracy comparaison}
\end{table}

\subsection{Multiclass Classification}
Since our objective is to identify the exact type of drone concerned in addition to the attack or normal cases, we performed the multiclass classification by designing tuples to represent the labels of our dataset. Initially, for the case of a four-class classifier, we employed a concise two-bit tuple format, denoted as $(a, b)$, where $a$ represents the UAV type and $b$ represents the normal or attack case. Thus, we have four cases as follows:\\
(0, 0) : ``DBPower - Normal"\quad \quad(0, 1) : ``DBPower - Attack"\\
(1, 0) : ``Parrot - Normal"\quad \quad \quad (1, 1) : ``Parrot - Attack". 
\\
Building upon this labeling, we extended our classification scope to encompass six classes. In this extension, we adopted a three-bit tuple representation, introducing labels such as (0, 0, 0), (1, 0, 0), (0, 1, 0), (0, 1, 1), (1, 1, 0), and (1, 1, 1). This strategy equipped our model to discern nuanced patterns in the dataset, enhancing our overall classification capabilities.
 \indent Furthermore, given the lack of data for the DJI Spark type in UF as shown in Table \ref{tab: dataset distribution}, we have kept only the two types of drone, Parrot Bebop and DBPower to build the combined dataset for the multiclass classification task of UF mode. 
 \begin{figure}[t]
 \centering
     \begin{subfigure}[b]{0.45\textwidth}
         \centering
    \includegraphics[width=\textwidth]{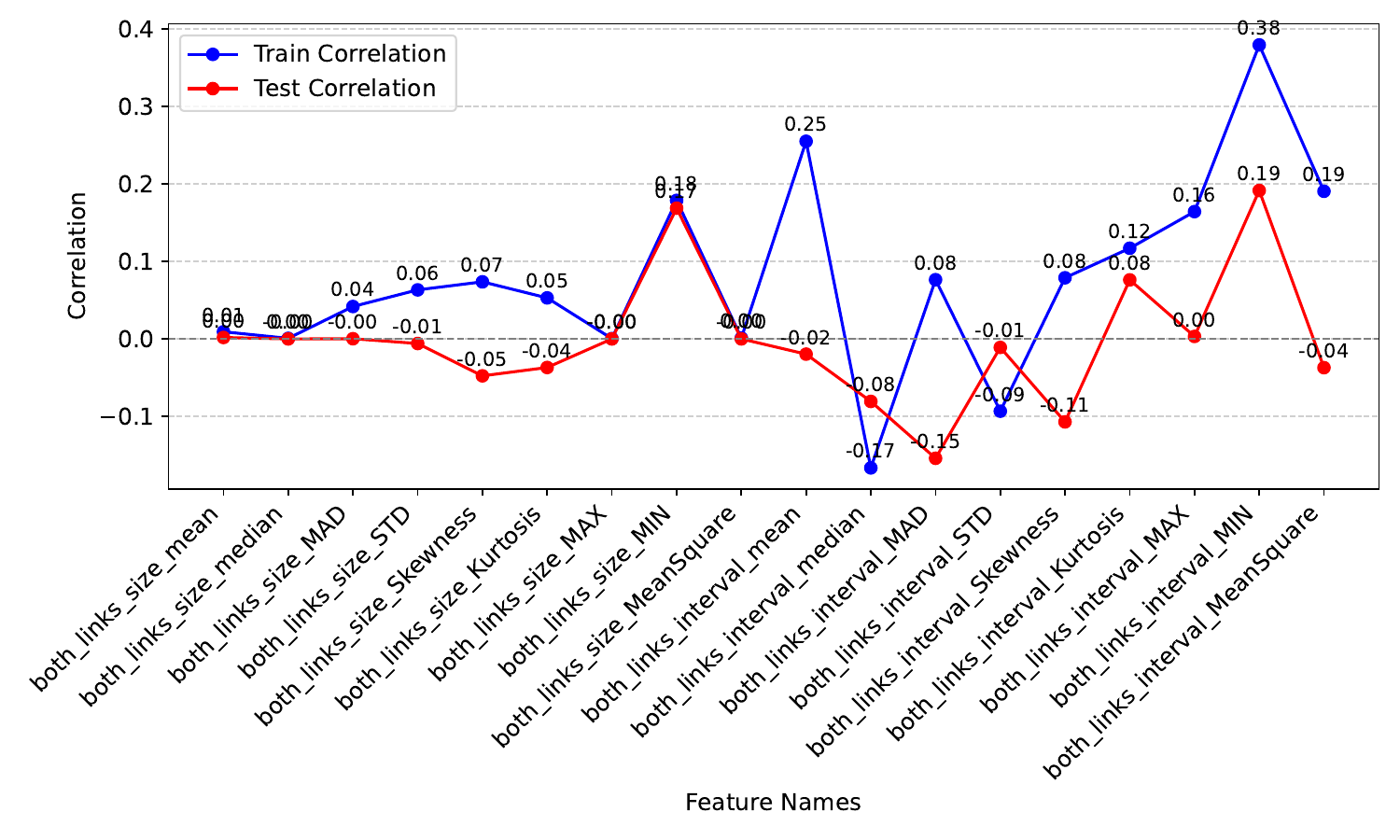}
    \caption{Correlation UF.}

    \label{fig: Correlation UF}
     \end{subfigure}
 
     \begin{subfigure}[b]{0.45\textwidth}
    \centering
    \includegraphics[width=\textwidth]{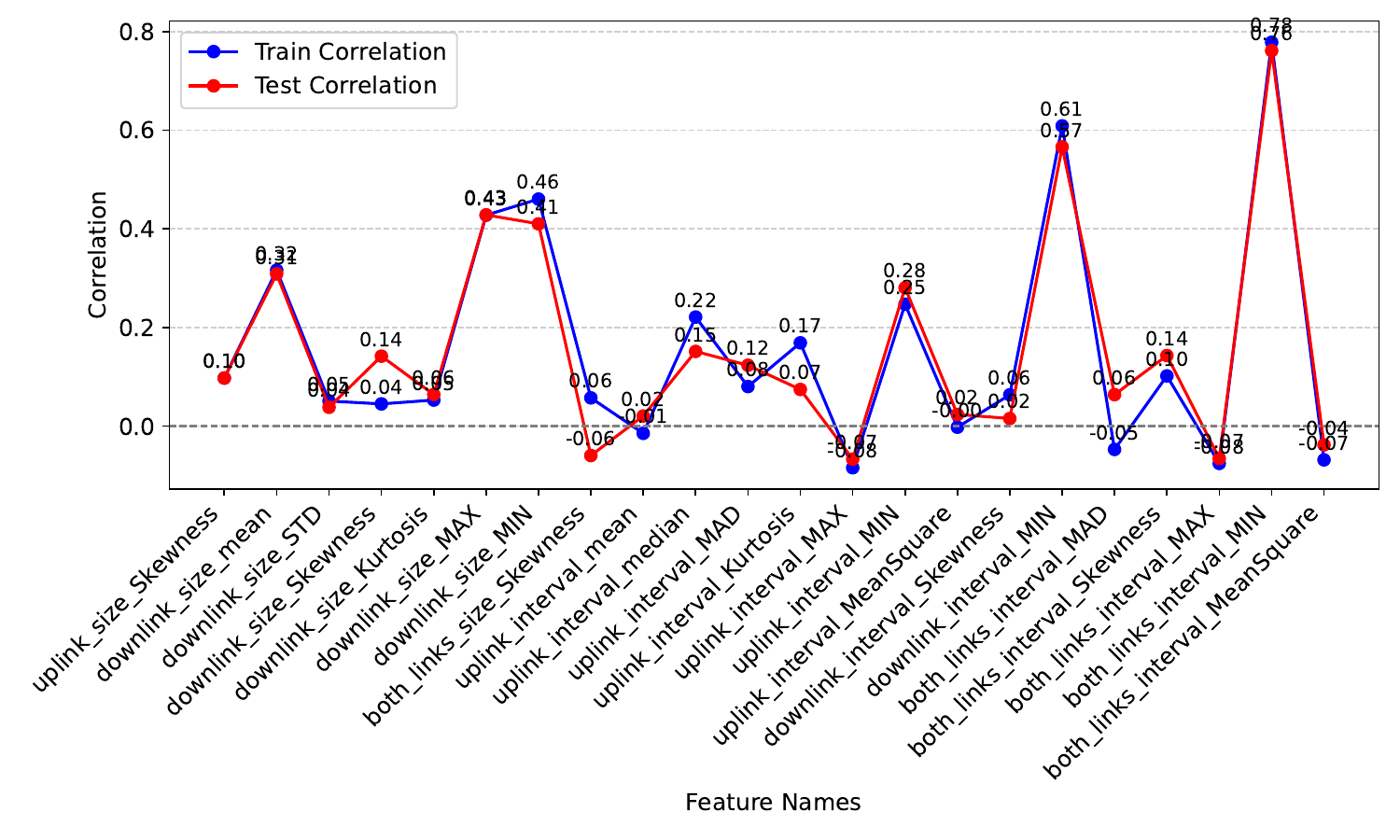}
    \caption{Correlation BF.}
    \label{fig: Correlation BF}
     \end{subfigure}
        \caption{Correlation between common features of the train and test datasets for bidirectional and unidirectional flow modes.}
        \label{fig: Correlation between common features of the train and test datasets for BF and UF.}
        \vspace{-1.8em}
\end{figure}

After collecting the datasets of both Parrot and DBPower types, we observed that these two UAVs generate data having similarities in some of their features. This behavior misleads our CNN in the multiclass classification to distinguish between the type of UAVs. To overcome this issue, we have calculated the cross-correlation between features of both UAVs, and we have removed features having a high correlation. Fig. \ref{fig: Correlation UF} represents the correlation between common features of the train and test datasets for UF mode. For example, features such as $both\_links\_interval\_MIN$, $both\_links\_interval\_mean$ were dropped from both the train and test datasets.

\begin{figure}[t]
    \vspace{-1em} 
    \centering
    \includegraphics[scale=0.365]{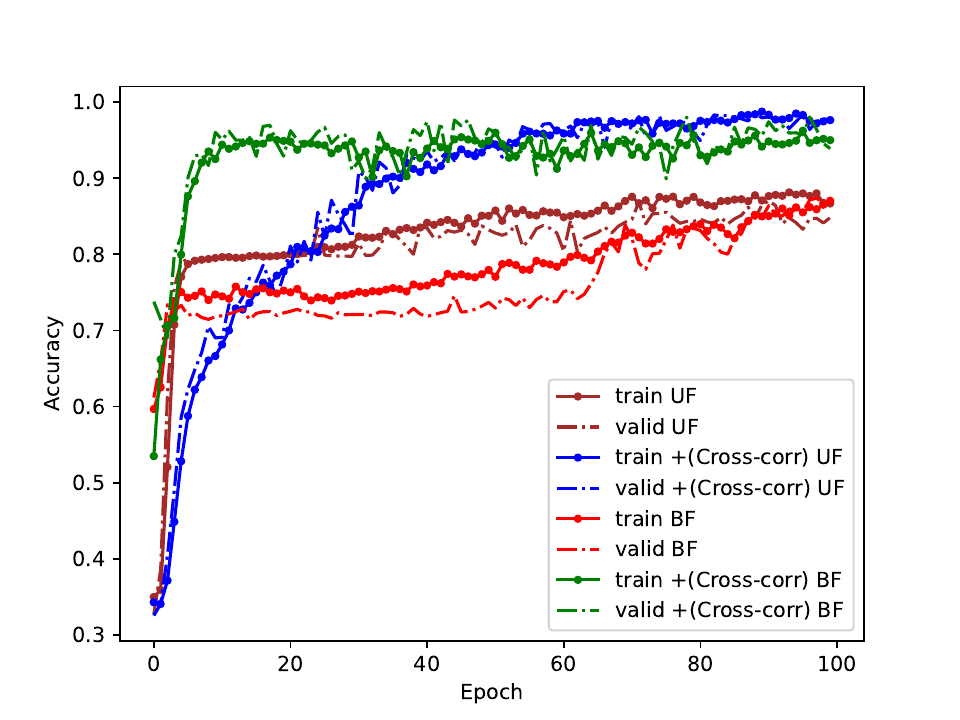}
\caption{Performance comparison of the proposed model with and without performing cross-correlation for both flow modes.}
    \label{fig: Comparaison CourbeAccuracy With And Without Correlation}
    \vspace{-0.5em} 
\end{figure}

For the sake of clarity, in Fig. \ref{fig: Correlation BF} we have shown only common features that are highly correlated of the train and test dataset for BF mode. As we can see from Fig. \ref{fig: Comparaison CourbeAccuracy With And Without Correlation}, when we perform the cross-correlation, the proposed model demonstrates significant improvement with an increase of about 10\% in accuracy compared to the case without cross-correlation. After performing the cross-correlation between the different datasets of the UAVs, a concatenated dataset from the previous datasets is constructed.

\begin{figure}[t]
 \centering
     \begin{subfigure}[b]{0.34\textwidth}
         \centering
         \includegraphics[width=\textwidth]{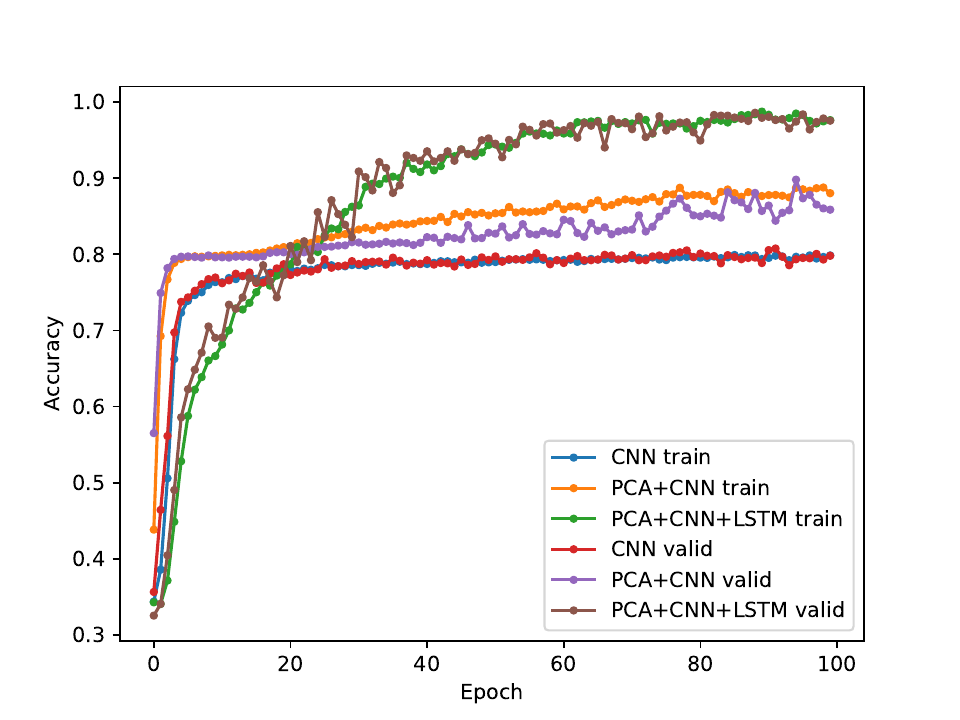}
         \caption{Average accuracy for UF.}
         \label{fig: Accuracy of MC UF}     
     \end{subfigure}
     \begin{subfigure}[b]{0.34\textwidth}
         \centering
         \includegraphics[width=\textwidth]{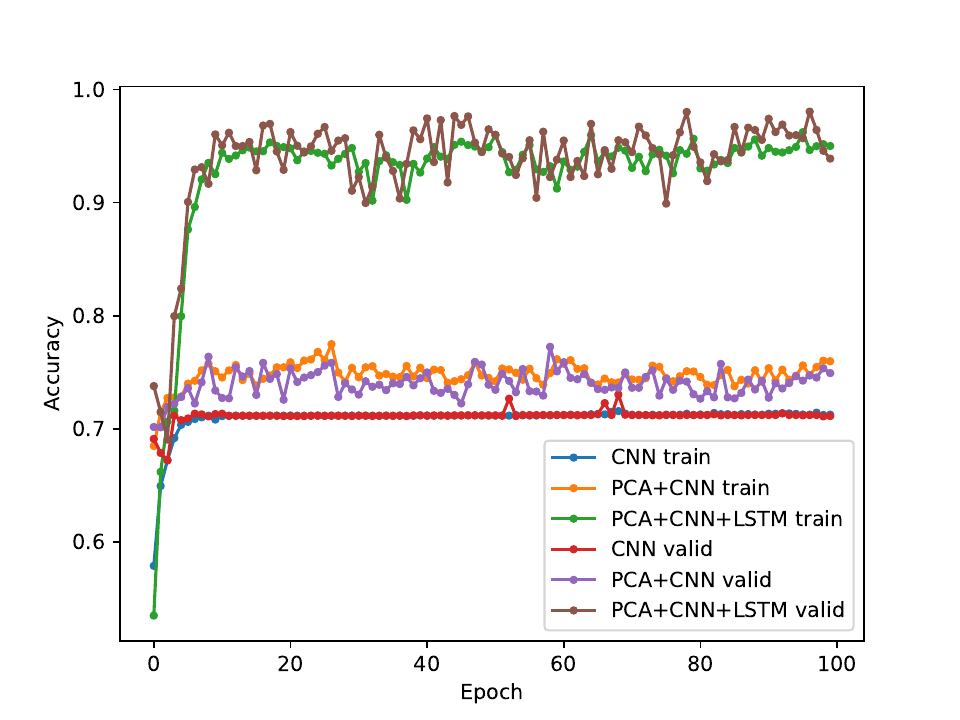}
         \caption{Average accuracy for BF.}
       \label{fig: Accuracy of MC BF}
     \end{subfigure}
        \caption{Performance comparison of the 1D CNN, PCA-1D CNN and PCA-1D CNN-LSTM for UF and BF.}
        \vspace{-1.5em}
        \label{fig:Accuracy of multiclasses UF and BF}
\end{figure}

Afterwards, the PCA method was performed on the combined dataset as shown in Fig. \ref{fig: Preprocessing Diagram} in order to reduce the concatenated dataset size. Therefore, the twelve components were zero-padded to make the input size of the model equal to 6x6, which provided better results. Finally, the model was trained and evaluated, and the results of the average accuracy for the UF and BF modes for the four-class classifier are presented in Fig. \ref{fig:Accuracy of multiclasses UF and BF}.\\
The obtained results allow us to conclude that our proposed model 1D CNN-LSTM enhanced with the PCA method and cross-correlation achieves good accuracy percentages up to 95\% for both communication flow modes, as shown in Fig. \ref{fig: Accuracy of MC UF} and Fig. \ref{fig: Accuracy of MC BF}. This shows that our model is better compared to the 1D CNN model and the 1D CNN model with the PCA method, which only achieve an accuracy around 75\% for the BF mode and ranging from 80\% to 90\% for the UF mode. 

\begin{table}[t]
\centering
  \caption{Comparison of model accuracy with existing solutions on the UAV-IDS-2020 dataset (\%)}
    \centering
    \begin{tabular}{|@{}c@{}|c@{} c@{} c|c@{} c@{ }c|}
        \hline
          \multirow{2}{*}{Model} & \multicolumn{3}{c|}{\textbf{BF}} & \multicolumn{3}{c|}{\textbf{UF}} \\
          \cline{2-7}
          &\resizebox{0.037\textwidth}{!}{Parrot }&\resizebox{0.037\textwidth}{!}{Power }&\resizebox{0.04\textwidth}{!}{ Comb. }& \resizebox{0.037\textwidth}{!}{Parrot }&\resizebox{0.037\textwidth}{!}{Power } & \resizebox{0.04\textwidth}{!}{ Comb. }\\
       \hline
           \resizebox{0.14\textwidth}{!}{ UAV-IDS-ConvNet \cite{10.1007/s00521-022-07015-9}  } & 99.99 & 99.99 & 99.98 & 99.58 & 82.78 & 84.20\\
          \hline
       
          \resizebox{0.07\textwidth}{!}{\textbf{Our model}} & \textbf{ 99.99 } & \textbf{ 99.99 } & \textbf{ 99.99 } & \textbf{ 99.99 } & \textbf{ 99.90 } & \textbf{99.98 }\\
       \hline
    \end{tabular}
    \label{tab: Comparaison Model}
    \vspace{-1.5em}
\end{table} 
 
\begin{figure}[t]
   \centering
    \includegraphics[scale=0.21]{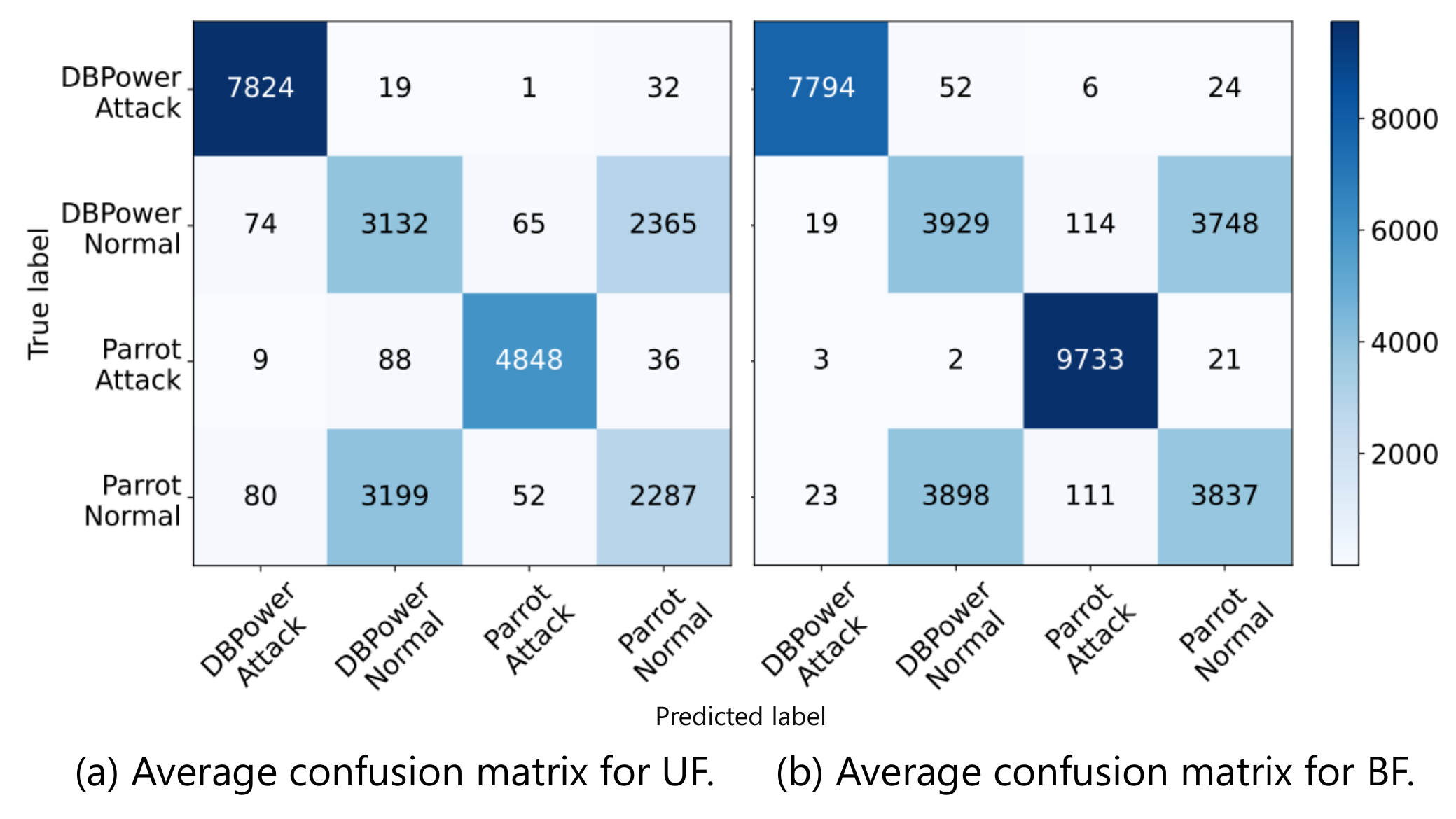}
     \caption{Four-class average classification performance for both flow modes.}
      \label{fig: Confusion matrix of MC for the BF and UF}
\end{figure}

\begin{figure}[t]
   \centering
    \includegraphics[scale=0.41]{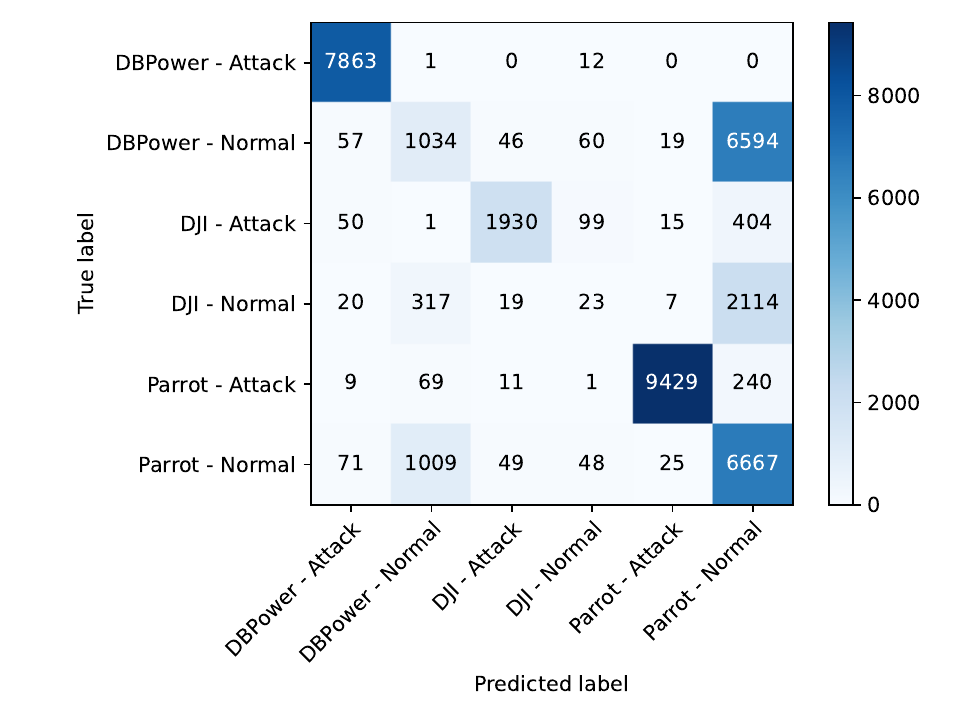}
        \vspace{-0.5em} 
\caption{Six-class average classification performance for BF.}
    \label{fig: Confusion matrix of MC of 6 classes for the BF}
    \vspace{-1.5em} 
\end{figure}

 On the other hand, from Fig. \ref{fig: Confusion matrix of MC for the BF and UF} representing the average confusion matrix of BF and UF modes for the four-class classification case, it can be seen that the model exhibits strong performance in distinguishing between attack and normal cases of unseen data, as indicated by the high accuracy along the diagonal elements of the confusion matrices. While in normal cases, the model may occasionally misclassify UAV types between DBPower-Normal and Parrot-Normal. However, in attack cases, it accurately identifies the type of UAV involved.\\
\indent After evaluating the four-class classifier, we extended our analysis to six classes. Despite the increased complexity, our model proved highly accurate in distinguishing between attack cases of the three types of UAVs. From the confusion matrix represented in Fig. \ref{fig: Confusion matrix of MC of 6 classes for the BF}, it's clear that our model excels at discerning attack cases, achieving an accuracy of 95\%.

\section{Conclusion}
In this paper, we proposed a new IDS for the detection and classification of drones. In fact, by using binary-tuple representation to encode class labels, and performing the cross-correlation along with \gls{PCA} method, we have enhanced the IDS performance in achieving accurate results.
The proposed system was trained and tested using the recent UAV-IDS-2020 dataset, where the experimental results showed a high potential of the proposed CNN-LSTM based approach in both binary and multiclass classification tasks. In our future work, we plan to investigate advanced data augmentation techniques to create more samples for the DJI Spark type. This will provide a complementary dataset that will enable to investigate further scenarios.

\section*{Acknowledgment}
This work was sponsored by the JFD program under the UM6P-EPFL Excellence in Africa Initiative.
\bibliographystyle{IEEEtran}
\bibliography{biblio_traps_dynamics}

\begin{thebibliography}{10}
\providecommand{\url}[1]{#1}
\csname url@samestyle\endcsname
\providecommand{\newblock}{\relax}
\providecommand{\bibinfo}[2]{#2}
\providecommand{\BIBentrySTDinterwordspacing}{\spaceskip=0pt\relax}
\providecommand{\BIBentryALTinterwordstretchfactor}{4}
\providecommand{\BIBentryALTinterwordspacing}{\spaceskip=\fontdimen2\font plus
\BIBentryALTinterwordstretchfactor\fontdimen3\font minus \fontdimen4\font\relax}
\providecommand{\BIBforeignlanguage}[2]{{%
\expandafter\ifx\csname l@#1\endcsname\relax
\typeout{** WARNING: IEEEtran.bst: No hyphenation pattern has been}%
\typeout{** loaded for the language `#1'. Using the pattern for}%
\typeout{** the default language instead.}%
\else
\language=\csname l@#1\endcsname
\fi
#2}}
\providecommand{\BIBdecl}{\relax}
\BIBdecl
\renewcommand{\BIBentryALTinterwordstretchfactor}{4}

\bibitem{Applications}
M.~Ghamari, P.~Rangel, M.~Mehrubeoglu, G.S. Tewolde, and R.S. Sherratt, ``Unmanned aerial vehicle communications for civil applications: A review,'' \emph{IEEE Access}, vol.~10, pp. 102\,492--102\,531, 2022.

\bibitem{choudhary2018intrusion}
G.~Choudhary, V.~Sharma, I.~You, K.~Yim, I.R. Chen, and J.H. Cho, ``Intrusion detection systems for networked unmanned aerial vehicles: A survey,'' in \emph{14th International Wireless Communications \& Mobile Computing Conference}, pp. 560-565, 2018.

\bibitem{10000726}
H.~Benaddi, M.~Jouhari, K.~Ibrahimi, A.~Benslimane, and E.M. Amhoud, ``Adversarial attacks against {I}o{T} networks using conditional {GAN} based learning,'' in \emph{IEEE Global Com. Conference}, pp. 2788-2793, 2022.

\bibitem{s22218085}
H.~Benaddi, M.~Jouhari, K.~Ibrahimi, J.~Ben~Othman, and E.M. Amhoud, ``Anomaly detection in industrial {I}o{T} using distributional reinforcement learning and generative adversarial networks,'' \emph{Sensors}, vol.~22, 2022.

\bibitem{8102043}
I.~G{\"u}ven\c{c}, O.~Ozdemir, Y.~Yapici, H.~Mehrpouyan, and D.~Matolak, ``Detection, localization, and tracking of unauthorized {UAV} and jammers,'' in \emph{36th IEEE/AIAA Digital Avionics Systems Conference}, pp. 1-10, 2017.

\bibitem{8556480}
I.~Bisio, C.~Garibotto, F.~Lavagetto, A.~Sciarrone, and S.~Zappatore, ``Blind detection: Advanced techniques for {W}i{F}i-based drone surveillance,'' \emph{IEEE Transactions on Vehicular Technology}, vol.~68, no.~1, pp. 938--946, 2019.

\bibitem{8846214}
B.~Taha and A.~Shoufan, ``Machine learning-based drone detection and classification: State-of-the-art in research,'' \emph{IEEE Access}, vol.~7, pp. 138\,669--138\,682, 2019.

\bibitem{ALSAD201986}
M.F. Al-Sa’d, A.~Al-Ali, A.~Mohamed, T.~Khattab, and A.~Erbad, ``{RF}-based drone detection and identification using deep learning approaches: An initiative towards a large open source drone database,'' \emph{Future Generation Computer Systems}, vol. 100, pp. 86--97, 2019.

\bibitem{9089489}
S.~Al-Emadi and F.~Al-Senaid, ``Drone detection approach based on radio-frequency using convolutional neural network,'' \emph{IEEE Inter. Conf. on Informatics, IoT, and Enabling Technologies}, pp. 29--34, 2020.

\bibitem{10.1007/s00521-022-07015-9}
Q.~Abu Al-Haija and A.~Al~Badawi, ``High-performance intrusion detection system for networked {UAV}s via deep learning,'' \emph{Neural Computing and Applications}, vol.~34, no.~13, pp. 10\,885--10\,900, 2022.

\bibitem{10.1145/3219819.3220117}
L.~Zhao and A.F. \textit{et al.}, ``Prediction-time efficient classification using feature computational dependencies,'' in \emph{Proceedings of the 24th ACM SIGKDD Inter. Conf. on Knowledge Discovery \& Data Mining}, 2018.

\bibitem{9264568}
A.~Mishra, A.M.K. Cheng, and Y.~Zhang, ``Intrusion detection using principal component analysis and support vector machines,'' in \emph{16th IEEE International Conference on Control \& Automation}, pp. 907-912, 2020.

\end{thebibliography}
\end{document}